\documentclass[%
 reprint,
 superscriptaddress,
 twocolumn,
 amsmath,amssymb,
 aps,
 prb,
]{revtex4}

\usepackage{graphicx}
\usepackage{dcolumn}
\usepackage{bm}
\usepackage{hyperref}

\usepackage{siunitx}
\usepackage{physics}

\usepackage{color}

\newcommand{\PHDG}{^{\vphantom{\dagger}}}

\begin{document}

\preprint{APS/123-QED}

\title{L-hole Pockets of the Palladium {F}ermi Surface Revealed by Positron Annihilation Spectroscopy}

\author{Michael Sekania}
\affiliation{Institut f\"ur Physik, Martin-Luther Universit\"at Halle-Wittenberg, 06120 Halle/Saale, Germany}
\affiliation{Theoretical Physics III, Center for Electronic Correlations and Magnetism, Institute of Physics, University of Augsburg, 86135 Augsburg, Germany}
\affiliation{Andronikashvili Institute of Physics, Javakhishvili Tbilisi State University, 0177 Tbilisi, Georgia \\ and Faculty of Natural Sciences and Medicine, Ilia State University, 0162 Tbilisi, Georgia}

\author{Andreas \"Ostlin}
\affiliation{Theoretical Physics III, Center for Electronic Correlations and Magnetism, Institute of Physics, University of Augsburg, 86135 Augsburg, Germany}
\affiliation{Augsburg Center for Innovative Technologies, University of Augsburg, 86135 Augsburg, Germany}

\author{Wilhelm H. Appelt}
\affiliation{Theoretical Physics III, Center for Electronic Correlations and Magnetism, Institute of Physics, University of Augsburg, 86135 Augsburg, Germany}

\author{S.~B.~Dugdale}
\affiliation{H.~H.~Wills Physics Laboratory,
University of Bristol, Tyndall Avenue, Bristol, BS8 1TL, United Kingdom}

\author{Liviu Chioncel}
\affiliation{Theoretical Physics III, Center for Electronic Correlations and Magnetism, Institute of Physics, University of Augsburg, 86135 Augsburg, Germany}
\affiliation{Augsburg Center for Innovative Technologies, University of Augsburg, 86135 Augsburg, Germany}


\date{\today}

\begin{abstract}
Using the combined Density Functional and Dynamical Mean Field theory we study relativistic corrections to the Fermi surface of palladium. We find indeed that relativistic corrections create a small hole pockets at the $L$-symmetry points. Furthermore we show that the computed two dimensional Angular Correlation of Electron Positron Annihilation Radiation (the so called $2D$-ACAR) clearly demonstrates the existence of these $L$-hole pockets, which remains robust against electronic correlations. A $2D$-ACAR experiment should therefore provide the ``smoking-gun'' proof for the existence of the $L$-hole pockets in the palladium Fermi surface.
\end{abstract}

\pacs{Valid PACS appear here}
\maketitle

\sloppy

\section{\label{sec:Inro}Introduction}
 
Despite its partially filled $d$-shell, elemental palladium (Pd) is not ferromagnetic.  It belongs, however, to a class of  materials called a nearly ferromagnetic metals~\cite{mori.85,lo.ta.85}. 
At low temperatures, Pd possesses the highest density of states (DOS) at the Fermi energy ($E_{\mathrm{F}}$) among all transition metals. 
The Stoner criterion is almost fulfilled~\cite{le.ru.92} and therefore Pd is considered to be situated at the verge of ferromagnetism. 
Nearly ferromagnetic metals exhibits strong magnetic fluctuations which are believed to have a great impact on macroscopic quantities such as the heat capacity and magnetic susceptibility~\cite{mori.85}.

Early band structure calculations predicted that the Fermi surface consists of four sheets.
An electron sheet centered at the $\Gamma$-point, open-hole sheets and hole pockets centered at $X$, and very small hole pockets at $L$~\cite{ande.70,mu.fr.70}. Experimental de Haas-van Alphen (dHvA) studies~\cite{wi.ke.71} identified the $\Gamma$ and $X$-centered sheets but failed to capture the small $L$-pockets. 
The detailed examination of the data did not allowed for an adequate fit of the ellipsoid centered at $L$~\cite{wi.ke.71}. A possible explanation was provided by Andersen~\cite{ande.70} who noted the existence of a saddle point (shallow band) in the dispersion (band structure) near the $L$-point which leads to a large effective mass and low velocities, unusual for such a small size pockets.
Unlike dHvA studies, ultrasonic attenuation experiments  found indications of existence~\cite{br.ka.72} of $L$-hole pockets.
Subsequent analysis of dHvA measurements taken at higher fields and lower temperatures in combination with parametrization of the Fermi surface was able to observe signals also from the small $L$ hole pockets~\cite{dy.ca.81,jo.cr.84}.
Most of the theoretical studies modeling the Fermi surface of Pd use Density Functional Theory (DFT)~\cite{jo.gu.89,kohn.99,jone.15}. Electronic interaction effects are included through approximate exchange-correlation potentials such as the Local Density Approximation (LDA) or its gradient corrected version (GGA).
For paramagnetic correlated ($3d$ and $4f$) electron systems it is well known that LDA/GGA calculations fail to provide the correct ground state properties.
A quantitative theory for the explanation of the electronic structure and the physical properties of such systems has been consistently developed during the last two decades in the form of a combination of density functional theory and Dynamical-Mean Field Theory (DMFT)~\cite{me.vo.89,ge.ko.96,ko.vo.04,ko.sa.06}, which is generally referred to as LDA+DMFT~\cite{ko.sa.06,held.07}. 
In the LDA+DMFT scheme, the LDA provides the {\it ab initio}, material dependent input (orbitals and hopping parameters), while the DMFT solves the many-body problem for the local interactions. 
Previously~\cite{os.ap.16} we have applied the LDA+DMFT method to Pd and showed that the agreement between the experimentally determined and the theoretical lattice constant and bulk modulus is improved when correlation effects are included. Furthemore we found that correlations modify the Fermi surface around the neck at the $L$-point while small correlation effects are seen on the tube structures of the Fermi surface. 
For relatively weak Coulomb interactions the spectral functions, obtained within the LDA+DMFT and GW~\cite{sc.ko.06,ko.sc.07} methods, show no major differences. Although in our studies a perturbative DMFT impurity solver was employed in the LDA+DMFT computations, our results agree well also with those obtained using non-local many-body impurity solvers~\cite{sa.re.18}. Using the same LDA+DMFT method we computed the changes in the phonon dispersions and phonon density of states of Pd under a hydro-static pressure. 
We found that electronic correlations at finite temperatures strongly renormalize the phonon frequencies and are the source of the so-called Kohn anomaly~\cite{ap.os.20}.
The wave-vectors at which the Kohn anomaly arise are the nesting vectors of the Fermi surface which for Pd are the tube structures along $X-W$ direction. 
In contrast to DFT results, for which the phonon frequency remains essentially constant in a large temperature range, correlation effects reduce the restoring force of the ionic displacements at low temperatures, leading to a mode softening~\cite{ap.os.20}. 

In this paper we report results of the Two Photon Momentum Density [TPMD or $\rho^{2\gamma}({\bf p})$] emerging from the process of annihilation of electron-positron pairs in bulk Pd. 
The annihilation process of electron-positron pairs in solids is well understood~\cite{west.73,kars.05}: As a consequence of momentum and energy conservation electron-positron pairs decay predominantly into two $\gamma$-quanta (photons). The resulting $\rho^{2\gamma}({\bf p})$ carries valuable information about the electron momentum density sampled by the positron.
Early electron-positron annihilation studies in metals already observed a strong electron-positron attraction superposed on the many-body correlations between the electrons~\cite{be.co.50}.
Recently, the methodology to compute the spin-polarized two-dimensional angular correlation of annihilation radiation ($2D$-ACAR) in combination with LDA+DMFT was developed~\cite{ce.we.16}.   
This made it possible, for example, to experimentally pinpoint the strength of the local electronic interaction in Ni~\cite{ce.we.16}. 
In a related study we performed the full 3D-electron momentum density reconstruction from the measured $2D$-ACAR in vanadium which allowed us to identify specific signatures of the Fermi surface~\cite{we.be.17}.
The presence of the Fermi Surface (FS) is signaled by the cusps in $2D$-ACAR spectra projected along specific directions. By the Lock-Crisp-West (LCW) back-folding method the momentum densities may be represented within the first Brillouin zone and features of the Fermi surface can be identified. 

The paper is organized as follows: After the theoretical description of computational methods Sec.~\ref{sec:theor}, we describe in Sec.~\ref{sec:band_FS} the spectral function and the Fermi surface 
modeled by the LDA(+DMFT) calculations in which we compared relativistic effects by artificially tuning the speed of light. 
Sec.~\ref{sec:EMD_2D-ACAR} starts with a brief description of the method to compute the momentum densities and $2D$-ACAR spectra using Green's functions. In the corresponding subsections (Sec.~\ref{subsec:EMD_ACAR}) we discuss the changes in the electronic densities induced by the presence of positrons and compare the theoretical and experimental spectra along some particular direction (Sec. ~\ref{subsec:EMD_EXP}). The summary section (Sec.~\ref{sec:SUMM}) concludes the paper.   

\section{\label{sec:theor}Theoretical techniques}

Electronic structure calculations were performed with the Spin-Polarized Relativistic Korringa-Kohn-Rostoker (SPR-KKR) code~\cite{EKM11}. 
In the LDA computations the exchange-correlation potentials parametrized by Vosko, Wilk, and Nusair~\cite{VWN80} were employed. 
The experimental lattice parameter for face-centered cubic (fcc) Pd is $3.8907\,$\AA, and a BZ-mesh of ${57\times57\times57}$ was used throughout the calculations.

In order to study the possible correlation effects within the framework of LDA+DMFT~\cite{ko.sa.06,held.07}
one adds the multi-orbital on-site interaction term,
${\mathcal H}_{U} = \frac{1}{2}\sum_{{i \{m, \sigma \} }} U\PHDG_{mm'm''m'''}
c^{\dag}_{im\sigma}c^{\dag}_{im'\sigma'}c\PHDG_{im'''\sigma'}c\PHDG_{im''\sigma}$,
to the LDA Hamiltonian ${\mathcal H}_\text{LDA}$.
The corresponding many-body problem described by the total Hamiltonian
${\mathcal H}= {\mathcal H}_\text{LDA} + {\mathcal H}_{U} - {\mathcal H}_\text{DC}$  
is solved using the LDA+DMFT method, where ${\mathcal H}_\text{DC}$ serves to eliminate double counting of the
interactions already included in ${\mathcal H}_\text{LDA}$. Here, $c\PHDG_{im\sigma}$($c^\dagger_{im\sigma}$) destroys (creates) an electron with spin $\sigma$ on orbital $m$ at the site $i$. The Coulomb matrix elements $U_{mm'm''m'''}$  are expressed in the standard way~\cite{im.fu.98} in terms of three Kanamori parameters $U$, $U'$ and $J$.
Charge and self-energy self-consistent LDA+DMFT computations are performed using the KKR approach~\cite{mi.ch.05}. Contrary to the Hamiltonian formulation, the KKR implementation of the LDA+DMFT uses the concept of the multiple-scattering theory. Namely, the solution for the single-site problem includes the local self-energy of the many-body problem, and allows for
the evaluation of the scattering path-operators and the real-space DMFT corrected Green's function. 
The many-body self-energy is constructed using the spin-polarized $T$-matrix fluctuation exchange (SPT-FLEX)~\cite{li.ka.97,PKL05} impurity solver. 
This impurity solver is fully rotationally invariant even in the multi-orbital version and is reliable when the interaction strength is smaller than the bandwidth, a condition which is fulfilled in the case of Pd. 
The LDA+DMFT computations require the parametrization of the interaction matrix $U_{mm'm''m'''}$ in terms of the average local Coulomb $U$ and the exchange parameter $J$. 
The value of $U$ is sometimes used as a fitting parameter.
However, recent developments made it possible, in principle, to compute the dynamical electron-electron interaction matrix elements with a good accuracy~\cite{AIG+04}, but with substantial variations associated with the choice of the local orbitals~\cite{MA08}. Since the parameter $J$ is not affected by screening it can be calculated directly within the LDA and is approximately the same for all 3d elements, i.e $J\approx0.9$~eV.

\begin{figure*}[thbp]
  \includegraphics[width=\textwidth]{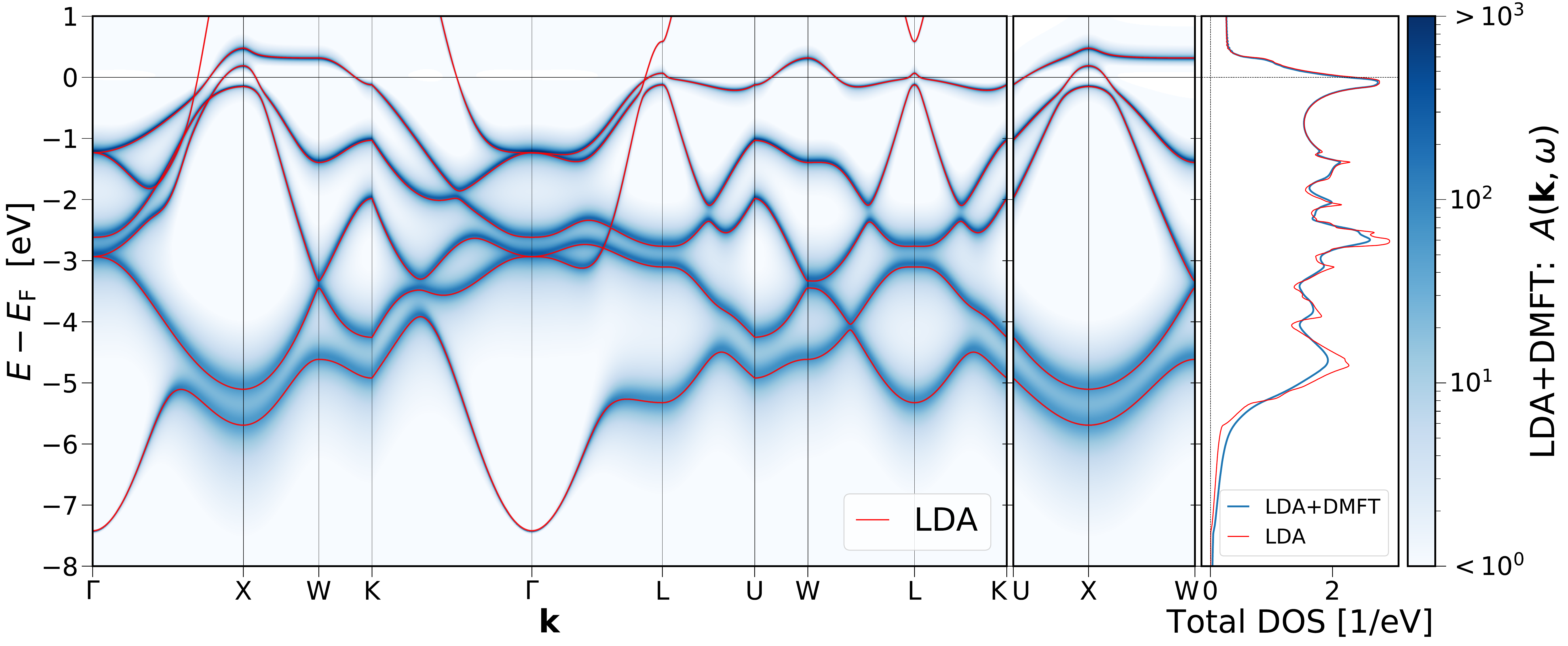}
  \caption{Relativistic band structure of Pd along high-symmetry directions in the BZ and the total density of states.
          }
  \label{fig:Fig1}
\end{figure*}

\section{Band structure and Fermi surface}
\label{sec:band_FS}
Fig.~\ref{fig:Fig1} presents the LDA+DMFT spectral function overlaid on the band structure. These results do not differ qualitatively from the previously reported results, see Ref.~\onlinecite{os.ap.16} and references therein. The Fermi level is situated near the top of the spectra in a relatively flat portion of the almost completely occupied $d$-bands. The density of states (DOS) is thus fairly large at the Fermi level, as is shown on the rightmost plot in Fig.~\ref{fig:Fig1}. Various details of the DOS have been recently discussed in Ref.~\cite{os.ap.16}. 
We focus our discussion on the hole pockets around the $L$-point which lies just above the Fermi energy. This particular band is very shallow, and as we move away from this specific point, it drops below the $E_{\mathrm{F}}$. Its existence was predicted by relativistic Augmented Plane Wave (RAPW) calculation~\cite{an.mc.68,ande.70,mu.fr.70}. The initial dHvA experimental results reported its absence~\cite{wi.ke.71}, the data reporting the position of the corresponding band as located below $E_{\mathrm{F}}$. Note that the $L$-pockets have small areas (extent in $k$-space) therefore according to the usual dHvA interpretation (that the angular variation of the effective mass should follow the angular variation in the area) holes in the $L$-pocket should have small effective masses. 
The fact that effective masses on the $L$-pockets are large reflects the non-parabolic nature of the bands that are actually very shallow. 
Later on more precise experiments~\cite{dy.ca.81,jo.ha.84,jo.cr.84} accompanied by theoretical parametrization of Fermi surface sheets together with various other experimental techniques such as magnetoacustics~\cite{br.ka.72} and PES~\cite{hi.ea.78} finally revealed its existence.  
Thus, from the gathered information at the $L$ symmetry points a hole-like closed surface is formed and because it lies near a band extremum, it should be nearly ellipsoidal in shape.

\begin{figure}[thbp]
\includegraphics[width=\linewidth]{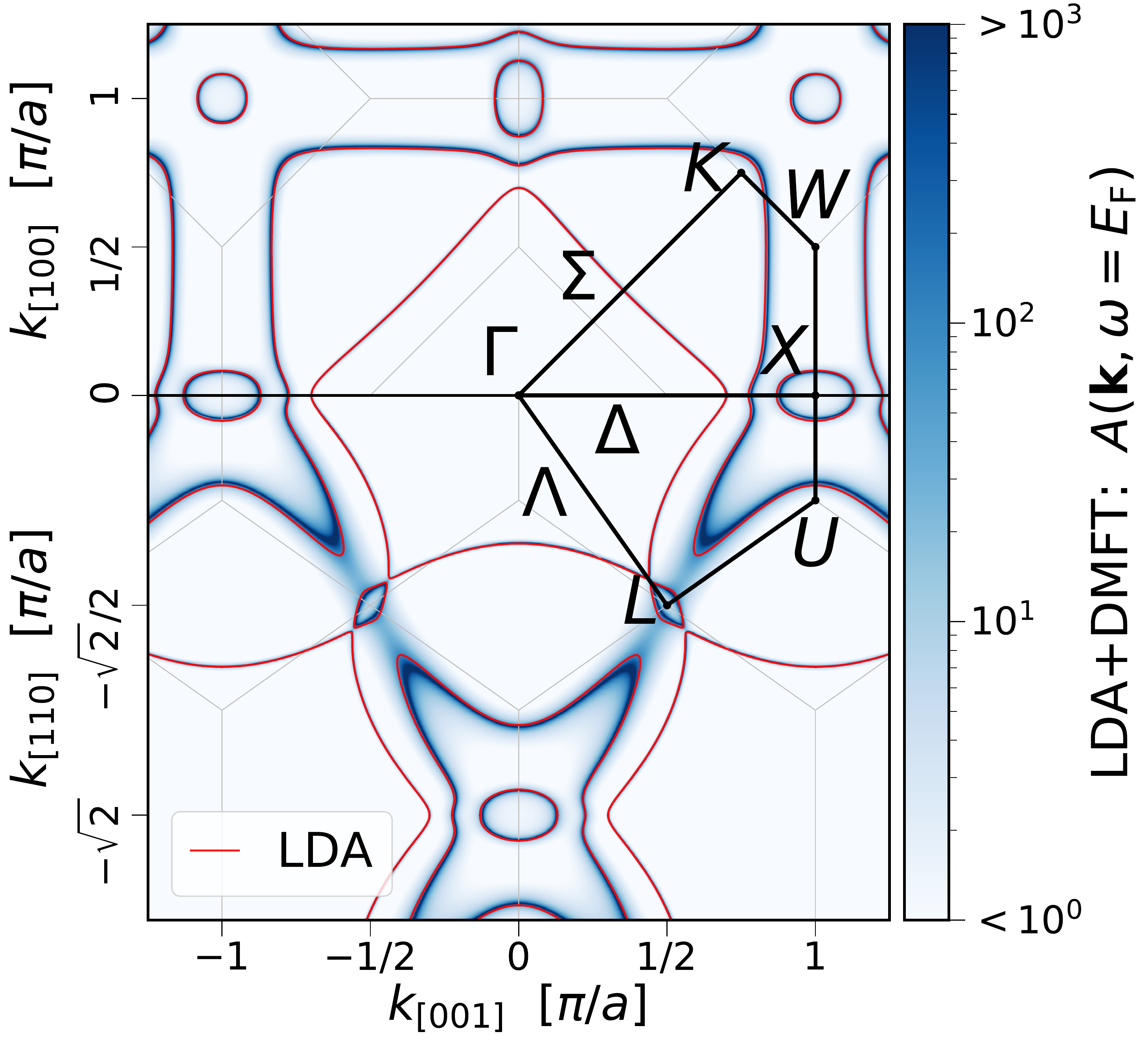}
  \caption{Fermi-surface cuts for ${k_{[010]}=0}$ (upper part) and ${k_{[1\bar{1}0]}}$ (lower part)
  planes.
          }
  \label{fig:Fig2}
\end{figure}

Another ellipsoidal hole pocket is found around the $X$-point. 
Along the ${\Gamma-X}$ direction its size is larger than along the ${X-W}$. 
At a slightly higher energies moving from $X$ to $W$ we follow a band situated above $E_{\mathrm{F}}$. This band passes below the Fermi level as we move form $W$ to $K$. 
The hole surface associated with this band is open in the ${X-W}$ direction forming the cylinders seen in Fig.~\ref{fig:Fig2}. The flatness of these bands makes them very sensitive to changes in the parameters.

The largest surface is that situated around the $\Gamma$-point. 
The volume of this surface is equal to the sum of the previously described hole
surfaces. This surface has extensions along the ${\langle 100 \rangle}$, ${\langle 110 \rangle}$, and ${\langle 111 \rangle}$ directions.
The open hole surfaces are particularly interesting as they are associated with large effective masses, and contribute significantly to the density of state at the Fermi level. 
As expected, the correlation effects broaden the one-particle bands, see Figs.~\ref{fig:Fig1}~and~\ref{fig:Fig2}, but nonetheless the overall shapes are not considerably changed with respect to the DFT-LDA results.

\begin{figure}[thbp]
  \includegraphics[width=\linewidth]{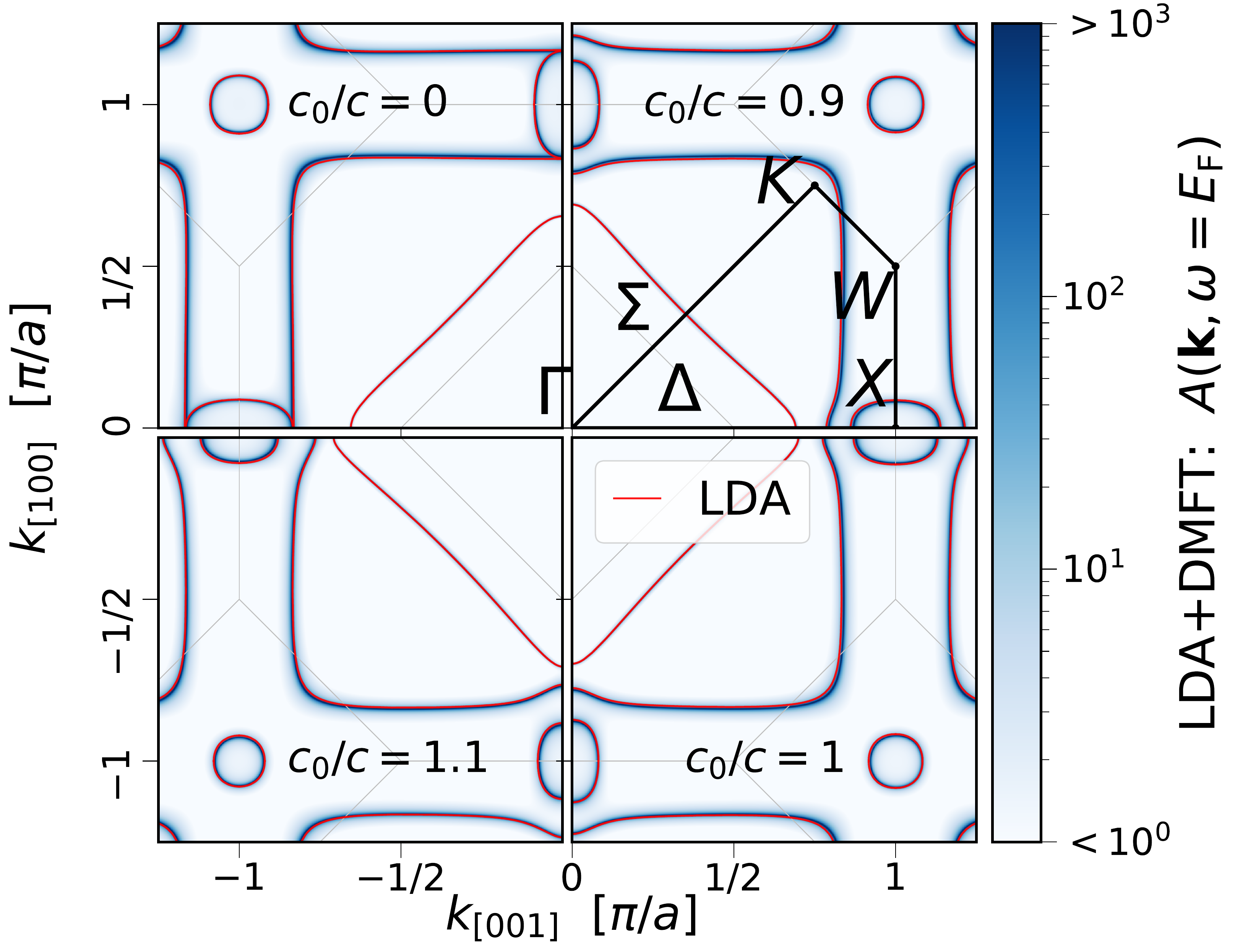}
  \includegraphics[width=\linewidth]{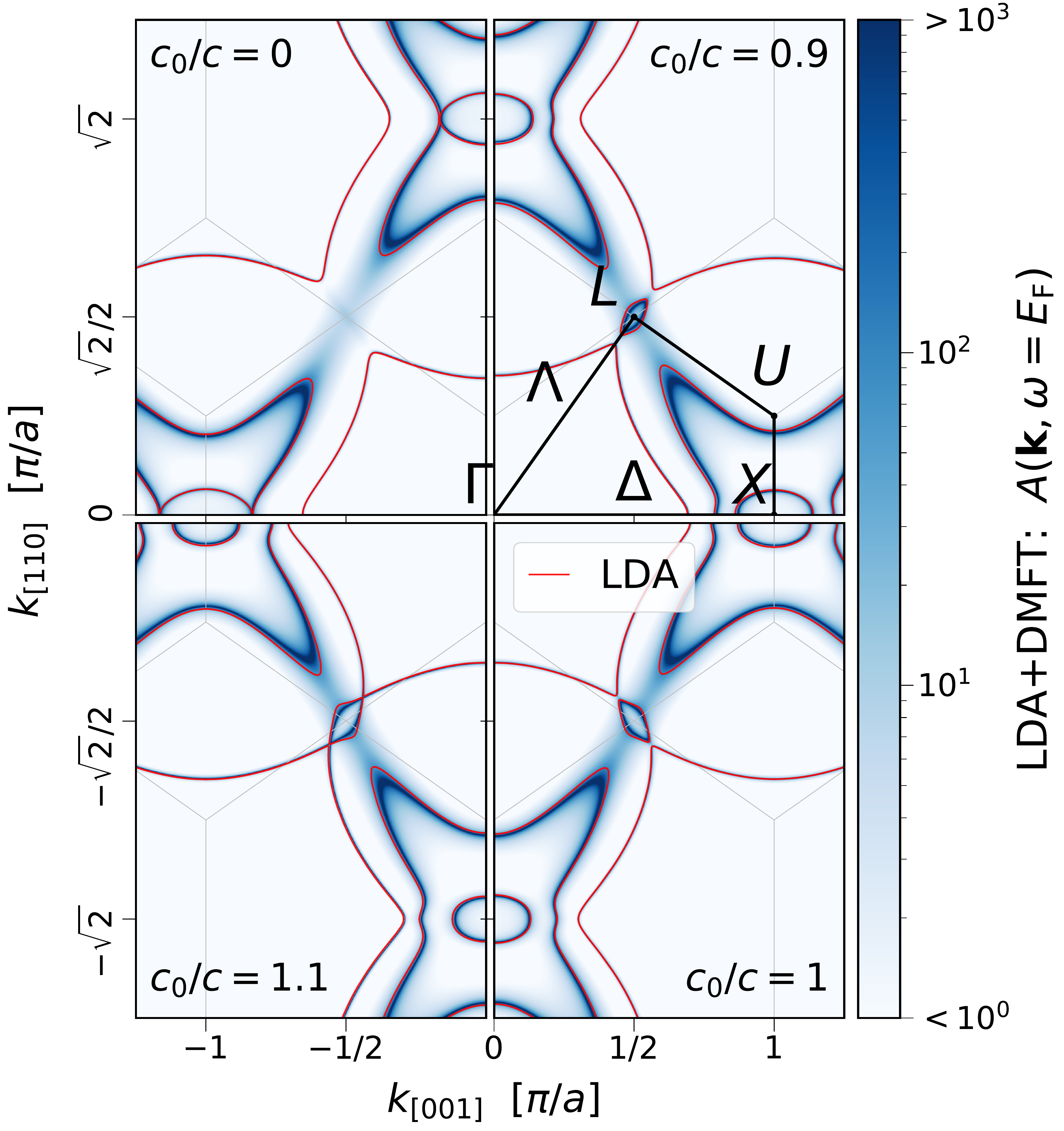}
  \caption{Fermi surface cuts for ${k_{[010]}=0}$ (upper plot) and ${k_{[1\bar{1}0]}=0}$ (lower plot)
  planes for different $c_0/c$ parameters.
          }
  \label{fig:Fig3}
\end{figure}

The change in the Pd Fermi surface cuts in the $(100)$ (${k_{[010]}=0}$) and $(1\bar{1}0)$ planes (${k_{[1\bar{1}0]}=0}$) are shown in Fig.~\ref{fig:Fig3} for various strengths of relativistic effects and in the presence of electronic correlations. 
Within the KKR approach the relativistic effects can be ``tuned'' by artificially modifying the magnitude of parameter describing the speed of light $c$ (the ratio $c_0/c$). 
Because the leading relativistic effects vary as $1/c^2$ they should all be affected accordingly.
Reducing the ratio $c_0/c$, i.e. increasing the speed of light, all relativistic effects are switched off and the description of the electronic structure is changed from the Dirac- to the Schr\"odinger-Kohn-Sham type equation. 
If on the other hand the speed of light is reduced, one enters into the so-called ``super-relativistic'' regime.
Calculations are performed within the "fully relativistic" setup with the exception $c_0=0$ which corresponds to the non-relativistic limit. 

For the non-relativistic case of (${c_0/c=0}$) no $L$-hole pockets are seen, while for all other ${c_0/c=0.9}$~and~${1.1}$ the Fermi surface consists of the same sheets with the same connectivity as was discussed in the previous LDA+DMFT computations~\cite{os.ap.16} (and references therein) with the scalar relativistic approach. 
Nevertheless, relativistic corrections change some details of the Fermi surface sheets.
The large ``jungle-gym'' hole surface (tubes) are enlarged (Fig.~\ref{fig:Fig3} upper panel) as well as the $\Gamma$ centered electron sheet, while the $X$ centered hole pockets are somewhat reduced in size (Fig.~\ref{fig:Fig3}). 
In the non-relativistic case at the $X$-point the ``jungle-gym'' and the hole pockets touch. For a non-zero value of the $c_0$ parameter these two sheets become distinct and their separation increases along with increasing $c_0$. As the magnitude of the spin-orbit interaction is proportional to a $1/c^2$ factor the separation of the ``jungle-gym'' and the hole pockets may also be associated to spin-orbit splitting.

Although the $L$ hole pockets are absent in the non-relativistic case (see the lower plot in Fig.~\ref{fig:Fig3}), it is interesting to note that the presence of electronic correlations effects (LDA+DMFT results) indicate the possible formation of this extra Fermi Surface pockets by the residual weight of the ``spikes'' (``jungle-gym'' spikes) along ${X-L}$ direction. Notably, the size of spikes is also increasing with increasing $c_0$.

\begin{table}[htbp]
 \center
 \caption{Fermi surface radii for $X$- and $L$-centered hole pockets (cf. Fig.~\ref{fig:Fig2}).
         }
 \label{tab:Tab1}
 \begin{ruledtabular}
 \begin{tabular}{ccc}
 \          & \multicolumn{1}{c}{LDA}          & \multicolumn{1}{c}{LDA+DMFT}    \\
 Directions & \multicolumn{1}{c}{($\pi / a$)}  & \multicolumn{1}{c}{($\pi / a$)} \\
 \colrule
 $X-\Gamma$ & $0.1290$  & $0.1240(5) $  \\
 $X-U$      & $0.0846$  & $0.0815(4) $  \\
 $X-W$      & $0.0821$  & $0.0791(4) $  \\
 \colrule
 $L-\Gamma$ & $0.0903$  & $0.081(1)\phantom{0}$  \\
 $L-U$      & $0.0398$  & $0.0317(5)$   \\
 $L-W$      & $0.0398$  & $0.0316(5)$   \\
 $L-K$      & $0.0399$  & $0.0317(8)$   \\
 \end{tabular}
 \end{ruledtabular}
\end{table}

To quantify the correlation effects on the Fermi surface hole pockets, we compared the radii of the $X$- and $L$-centered hole pockets and found a reduction of about $4\%$ for $X$- and about $20\%$ for $L$ in the linear dimensions of the Fermi sheets (see Table~\ref{tab:Tab1} and Fig.~\ref{fig:Fig1}).
In table~\ref{tab:Tab1} we list the symmetry plane radii of the $X$- and $L$-centered hole pockets.
The two columns correspond to the relativistic LDA and LDA+DMFT calculations.
Early theories to elucidate the role of electronic interactions were based on the paramagnon model~\cite{be.sc.66,do.en.66} and were focused on the effective mass enhancement computed from the self-energy which was taken as frequency dependent but not momentum dependent, i.e. retarded in time but local in space. 
Previously we have computed the quasi-particle weights ${Z=(1-\partial \Re[\Sigma(E)]/\partial E|_{E_{\mathrm{F}}})^{-1}}$ for the  different $U$ values
in the range from ${U=1-4}$~eV. The corresponding mass enhancement factor  was found to be in the range ${Z=0.975-0.916}$ which provides for the effective mass ratios ${m^*/m_{\text{LDA}}=Z^{-1}=1.03-1.09}$,  where $m_{\text{LDA}}$ is the LDA band mass. The experimental estimated value from the specific heat measurement is larger than the mass enhancement due to correlation effects even when the electron-phonon and the paramagnon contribution is added~\cite{os.ap.16}.  

\begin{figure}[t]
  \includegraphics[width=\linewidth]{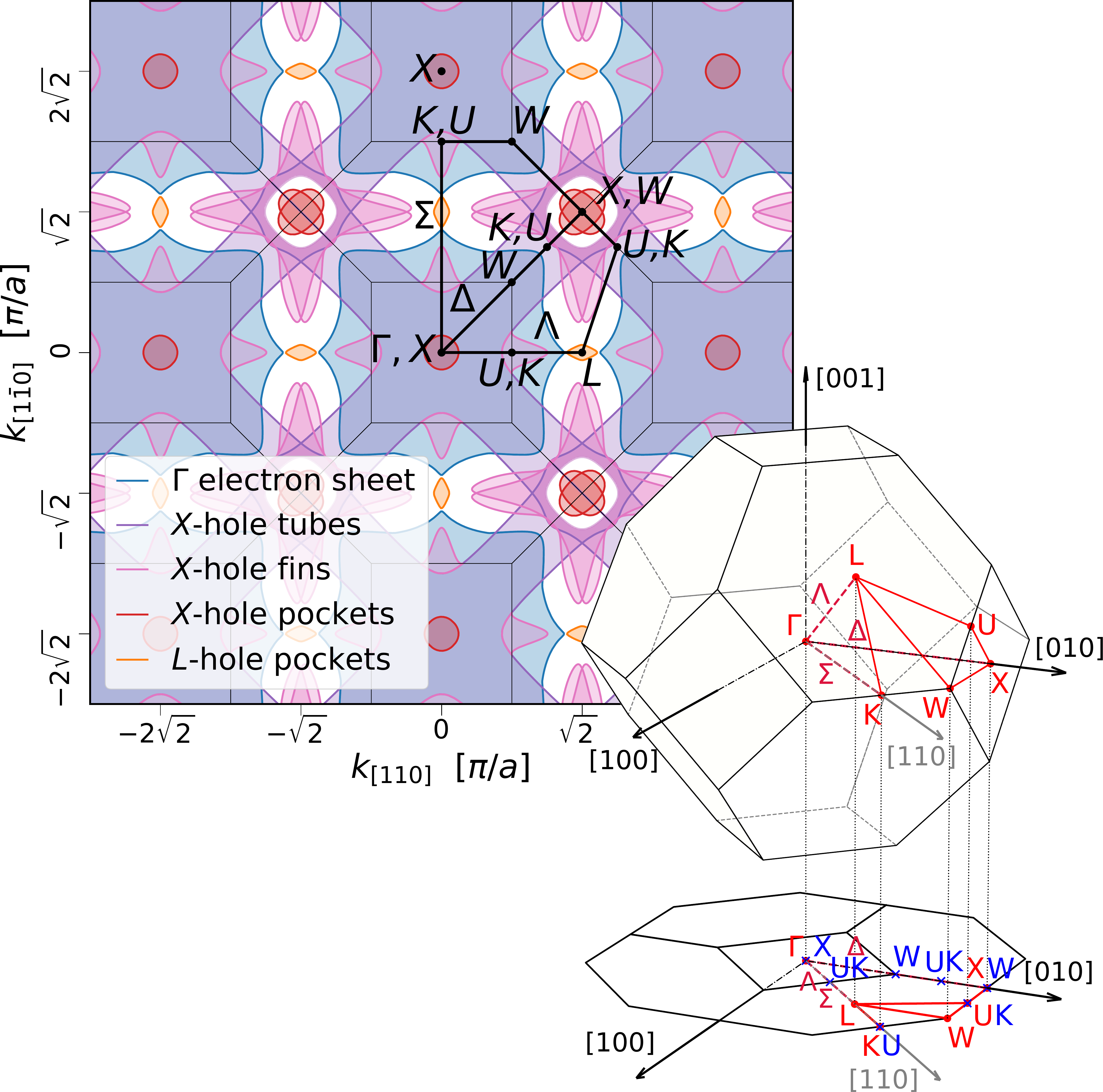}
  \caption{2D Fermi-surface projection along $[001]$ direction from LDA calculations (upper-left plot).           Normal regions are blank.
           $X$-hole fins shown here separately are part of the $X$-hole tubes.
           The bottom-right plot: the Brillouin zone for the fcc crystal structure (upper part) and $2D$ projection of it along the $[001]$ direction (lower part).
           High-symmetry points and directions, as well as extra images of high-symmetry points on the projection plane (blue crosses) are also shown.
           In the 2D FS projection (upper-left plot), the $\Sigma$-direction in ${\Gamma-K-W-X-\Gamma}$-path is along the $[1\bar{1}1]$ direction.
         }
  \label{fig:Fig4}
\end{figure}

In Fig.~\ref{fig:Fig4} we show the projected Fermi surface along $[001]$ direction.
Note that the small $L$-pockets that are situated on the hexagonal faces remain clearly visible also in the $2D$ projected FS, and are not overlapped by other FS sheets in the projection along this direction.
By the projection, the hole pockets centered at the $X$-point (square faces of fcc BZ) two out of the six, when projected down into the $\Gamma$-point, have a circular shape (cut across minor axis of revolution ellipsoid) with the remaining four projected into the $W$-point are identified as two crossing ellipses with major and minor axis exchanges (one of ellipses comes from the neighboring BZ).
The ``jungle-gym'' structure is also easily recognizable by (anti-)diagonals along projected the $\Gamma$-points. $X$-hole fins which are the part of the ``jungle-gym'' structure (shown as a separate FS sheet in Fig.~\ref{fig:Fig4}) can be identified as cross like features centered at $(X,W)$-points on projection plane. The ``jungle-gym'' tubes, perpendicular to the projection plane,  are identified around the $(W,X)$-points with the projected $X$-hole pockets inside.
Projections of the high-symmetry points $K$ and $U$ coincide with some of the $X$- and $W$-points.
In Fig.~\ref{fig:Fig4} we also show the fcc BZ with high-symmetry points and their projection on ${k_{[100]}=0}$ plane together with extra images of high-symmetry points (blue crosses and letters in projection plane).
This projected FS sheets image turns to be useful when considering the LCW back-folded EMD and $2D$-ACAR which will be discussed in the following sections.

\section{Electron momentum densities, $2D$-ACAR and Fermi surface}
\label{sec:EMD_2D-ACAR}
The theoretical analysis of the $2D$-ACAR spectra requires the knowledge of the two-particle electron-positron Green's function, describing the probability amplitude for an electron and a positron propagating between two different space-time points.
The DFT can be generalized to the problem at hand by including the positron density in the form of a 2-component DFT~\cite{bo.ni.86,pu.ni.94}. 
In the present calculations the electron-positron correlations are taken into account by a multiplicative (enhancement) factor, 
which results from the inclusion of the electron-positron interaction in the form of an effective one-particle potential as formulated by Boro\'nski and Nieminen~\cite{bo.ni.86} or alternatively by Drummond~\cite{dr.lo.11}.
Many-body DMFT corrections are taken into account only for the electronic subsystem and no many-body correlation is included for the positron. The Green's function of the electron-positron pair in momentum representation can be expressed using the multiple scattering Green's function and the eigenfunctions of the momentum operator $\phi_{{\bf p} \sigma'}$ as:
\begin{widetext}
\begin{align}
  G^\mathrm{X}_{\sigma \sigma'}({\bf p}_e,{\bf p}_p, E_e, E_p)
  = \frac{1}{N \Omega}
  \int d^3 {\bf r}
  \,
  \int d^3{\bf r^\prime}  \phi_{ {\bf p}_e \sigma}^{e\dagger}({\bf r})
  \Im G\PHDG_{e \sigma}({\bf r}, {\bf r^\prime}\,E_e)
  \phi_{{\bf p}_e \sigma}^{e}({\bf r^\prime})
  \,
  \phi_{{\bf p}_p \sigma^\prime}^{p\dagger}({\bf r})
  \Im G\PHDG_{p \sigma^\prime}({\bf r}, {\bf r^\prime},E_p)
  \phi_{{\bf p}_p \sigma'}^{p}({\bf r^\prime})\,,
  \end{align}
\end{widetext}
where ${\mathrm{X}=\text{LSDA(+DMFT)}}$ and $({\bf p}_e \sigma)$ and $({\bf p}_p \sigma^\prime)$
are the momenta and spin of electron and positron, respectively.
Taking into account the positron thermalization (which in metals occurs very quickly compared to the lifetime of the positron), one may
consider the positron to be in a state with ${{\bf p}_p=0}$ with $s$-type
symmetry and the energy corresponding to the bottom of the
positronic band.
The positron will almost always annihilate via
two photons from an anti-parallel positron-electron spin
configuration. As consequence, the electron-positron pair has zero
total spin angular momentum and ${\sigma=-\sigma'}$ holds.
The two-photon momentum density resulted from the annihilation process can be expressed
using the electron-positron \linebreak Green's function as
\begin{equation}\label{eq:rho}
  \rho^{2\gamma, \mathrm{X}}_{\sigma}({\bf p})=
  -\frac{1}{\pi}\!\int_{-\infty}^{\infty}\!\!\!\!\!\mathrm{d}E_e\!\! \int_{-\infty}^{\infty}\!\!\!\!\!\mathrm{d}E_p\,G^\mathrm{X}_{\sigma \sigma^\prime}({\bf p}_e,{\bf p}_p, E_e, E_p),
\end{equation}
whereas for the electron's momentum density (EMD) the electronic Green's function in the momentum representation is employed, 
\begin{equation}\label{eq:EMD}
  \rho^{\text{EMD}, \mathrm{X}}_{\sigma}({\bf p})=-\frac{1}{\pi}\int_{-\infty}^{\infty}\mathrm{d}E_e\, \Im G^\mathrm{X}_{\sigma \sigma^\prime}({\bf p}_e, E_e)\,.
\end{equation}
The energy integrations in Eqs.~\eqref{eq:rho}~and~\eqref{eq:EMD} are performed along a contour in the complex plane. 
For the two-photon momentum densities, in Eq.~\eqref{eq:rho}, integration over $E_p$ is not required as it is taken as a parameter corresponding to the bottom of the positronic band. 
Only the filled states contribute to the $2D$-ACAR spectra, as the momentum distributions are obtained through energy integrals up to the Fermi level. 
The computed $2D$-ACAR/EMD spectrum $\tilde{\rho}(p_x,p_y)$ is a $2D$ projection of the $3D$ momentum-density distribution ${\rho^{2\gamma/\text{EMD},\mathrm{X}}_\sigma({\bf p})}$ along a chosen $p_z$,
\begin{equation}\label{eq:2dacar}
  \tilde{\rho}^{2\gamma/\text{EMD},\mathrm{X}}_\sigma(p_x,p_y) = \int\mathrm{d}{p_z}\, \rho^{2\gamma/\text{EMD},\mathrm{X}}_\sigma({\bf p}) \,.
\end{equation}
In the following, the spin index $\sigma$ can be dropped, as the calculations are performed for the non-magnetic state. 

In order to expose details of the momentum densities the anisotropy spectrum $\rho_{an}({\bf p}_\perp)$ is obtained by subtracting an angular averaged contribution $\bar{\rho}(p)$.
The same procedure is usually used for the analysis of the experimental measurements,
which helps to emphasise the anisotropic features in the momentum density which arise from wavefunction anisotropy or, in a metal, from the presence of the Fermi surface.
All EMD and ACAR calculations were performed with cut off $|\mathbf{p}|_{\max}=16$~a.u. and step size $0.01$~a.u.

\begin{figure*}[t]
  \includegraphics[width=0.49\linewidth]{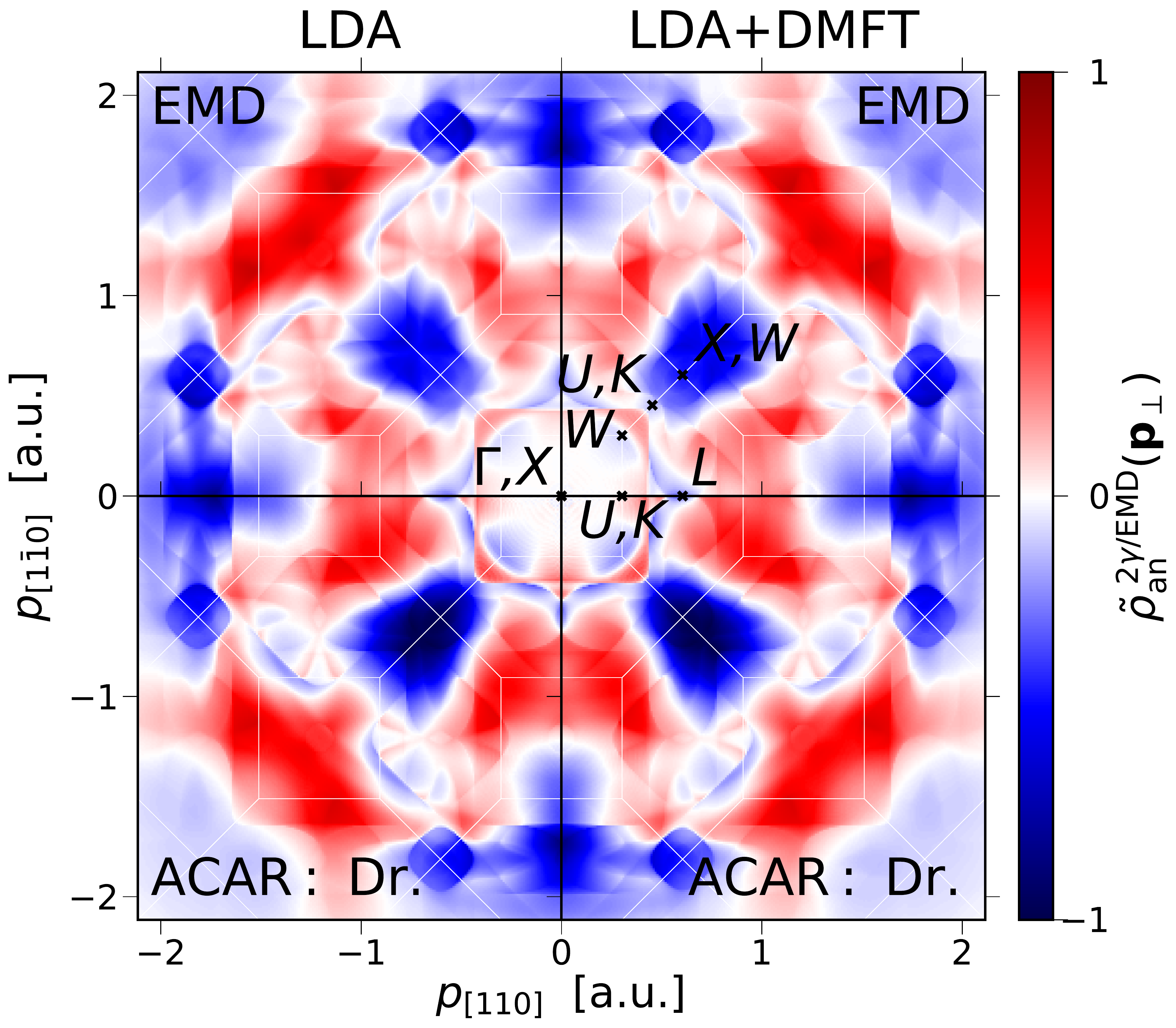}
  \hfill
  \includegraphics[width=0.49\linewidth]{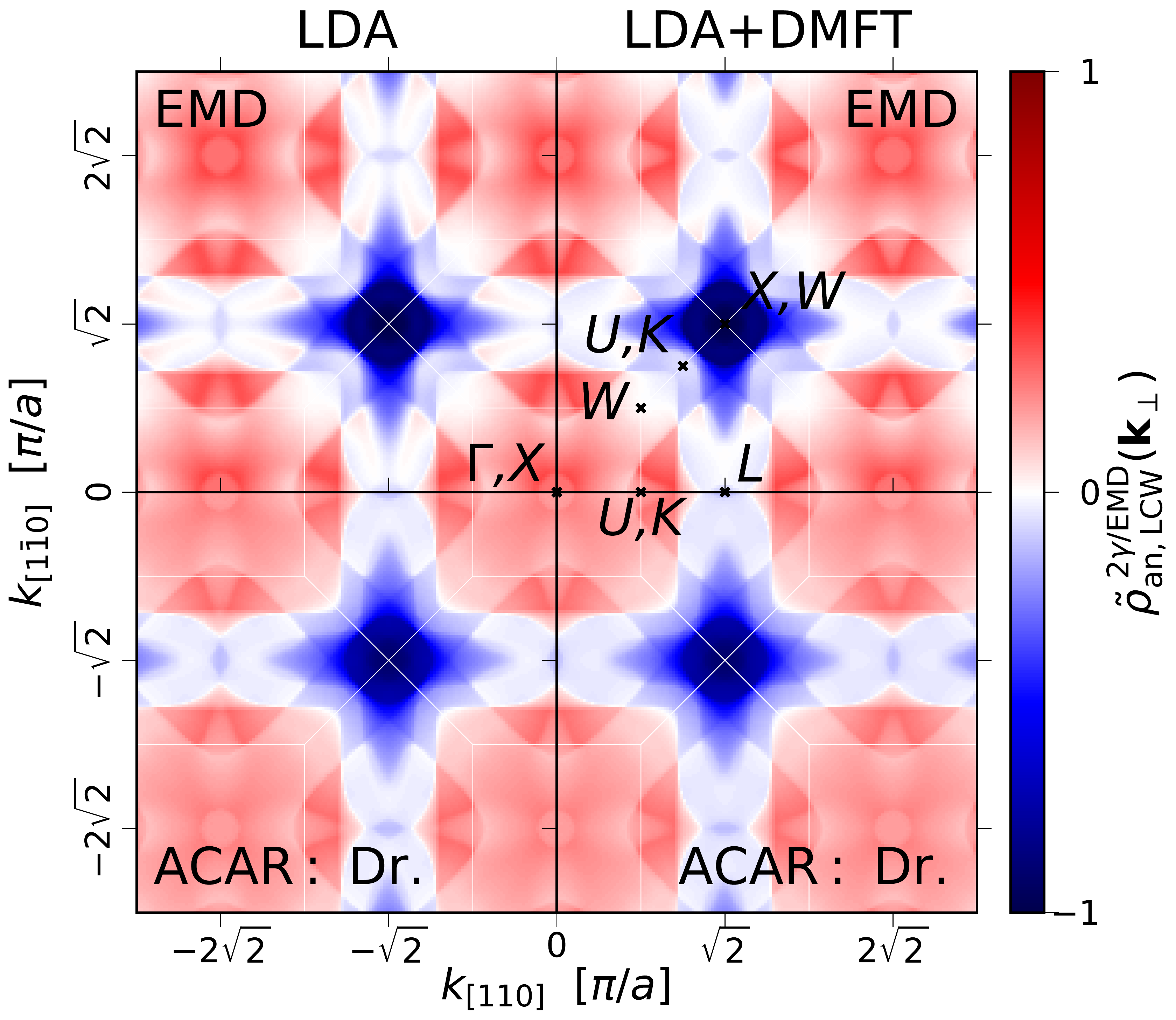}
  \caption{Anisotropy (left) and LCW back-folded anisotropy (right) plots of $2D$ projection of EMD (upper rows) and $2D$-ACAR spectra (bottom rows) from LDA (left column) and LDA+DMFT (${U=1.0}$~eV, ${J=0.3}$~eV) (right column) calculations.
           The $2D$ projection of EMD and $2D$-ACAR spectra are normalized by the number of valence electrons.
           The false-color map is normalized by maximal amplitude of anisotropy taken for all four cases.
           Projections of the high-symmetry points and directions (black crosses and lines), as well as extended BZ (white lines) are also shown.
          }
  \label{fig:Fig5}
\end{figure*}

We analyse the two-dimensional projections of the electronic momentum densities along the $[100]$ and $[111]$ directions (integration directions).
In the left plot in Fig.~\ref{fig:Fig5} we show the radial anisotropy of EMD (upper panels) and $2D$-ACAR spectra (lower panels) obtained with the LDA exchange-correlation (left column) and including dynamical correlations (right column), in the extended momentum space. 
Given that the effects of electronic correlations amount to slight difference between the spectral function and the one-particle band structure, see Fig.~\ref{fig:Fig1}, very similar EMD spectra are expected. Indeed we note that correlation effects does not change significantly the LDA EMD and $2D$-ACAR spectra.  
However, a direct comparison of the EMD and $2D$-ACAR spectra allows certain differences to be identified stemming form the presence of the positron in the computation. The most notable differences happen along the direction $(\Gamma,X)-W-(U,K)-(X,W)$ and in the second Brillouin zone around e.g., ${\mathbf{p}_\perp\approx(\pm1.5, \pm0.25)}$~a.u. points. Another visible effect happens in the region around the $W$-point in which the $2D$-ACAR spectra are depleted with respect to the EMD. In the region between $W$ and $(U,K)$ the $2D$-ACAR has a stronger signal, while further out following the same direction the spectra is loosing weight.
The ``jungle-gym'' structure in the ${X-W-X}$ direction remains visible in the EMD, also after the projection along $[001]$ which is because no intense momentum density exists around the $X$-point. In contrast, in the same region the electronic momentum density is enhanced by the presence of positrons as seen in the $2D$-ACAR spectra and the ``jungle-gym'' structure is covered by the response from the $X$-hole pockets. 
Strong positron enhancement effects are identified in the second Brillouin zone around e.g.: ${\mathbf{p}_\perp\approx(\pm1.5, \pm0.25)}$~a.u. points.
These might be used to asses the effects of electron-positron enhancement factor (in our case modeled by Drummond parametrization) if the experimental results would be available.

\begin{figure}[!hthp]
  \includegraphics[width=0.99\linewidth]{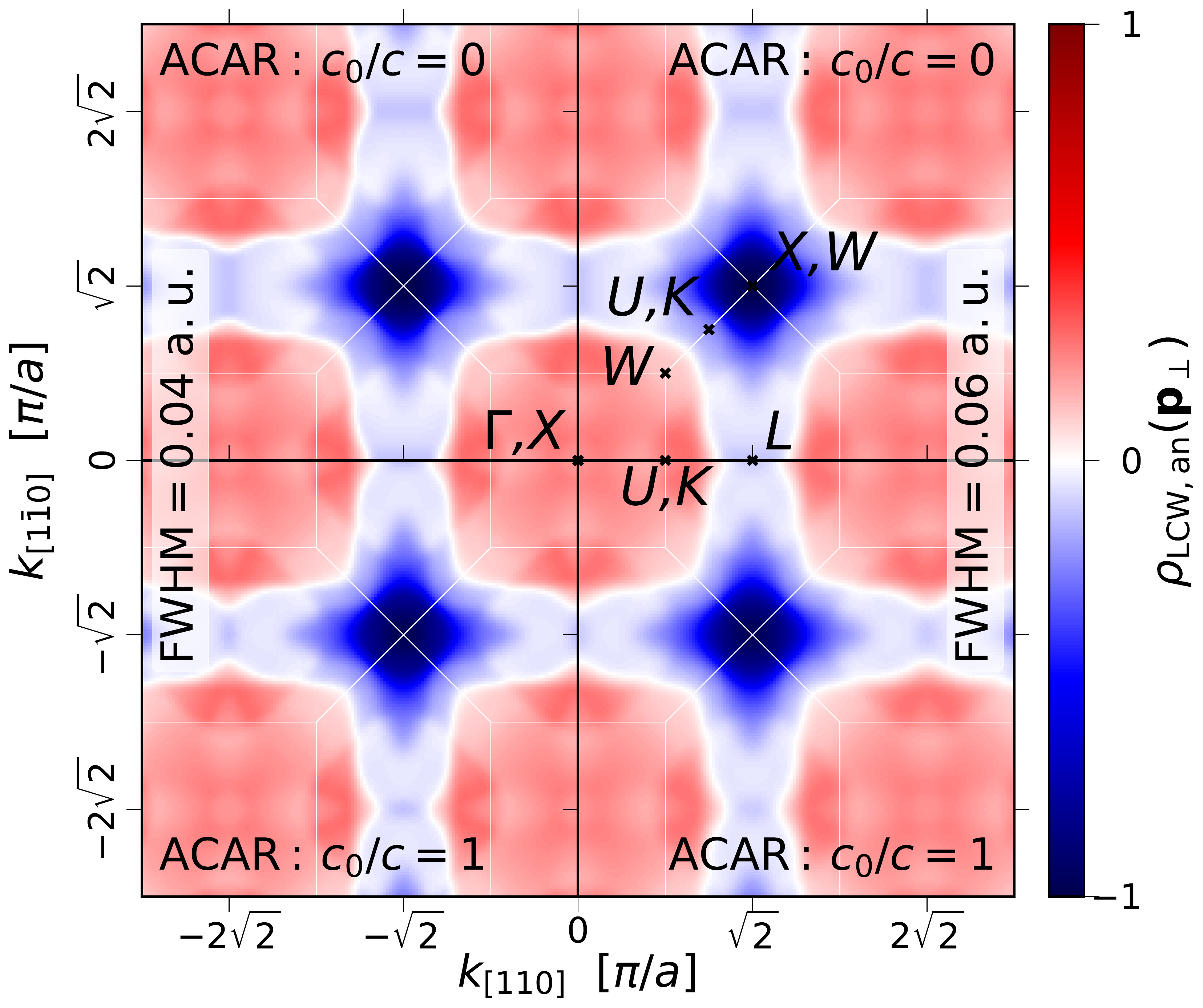}
  \caption{LCW back-folded anisotropy plots of $2D$-ACAR spectra with broadening of FWHM=$0.04$~a.u. (left column) and FWHM=$0.06$~a.u. (right column) from nonrelativistic (upper row) and relativistic (lower row) LDA+DMFT (${U=1.0}$~eV, ${J=0.3}$~eV) (right column) calculations.
           The $2D$-ACAR spectra are normalized by the number of valence electrons.
           The false-color map is normalized by maximal amplitude of anisotropy taken for all four cases.
           Projections of the high-symmetry points and directions (black crosses and lines), as well as extended BZ (white lines) are also shown.
          }
  \label{fig:Fig6}
\end{figure}

To compute the momentum densities in the corresponding $2D$-ACAR Brillouin-zones we employ the LCW back-folding procedure~\cite{lo.cr.73},
\begin{equation}\label{eq:LCW}
  \tilde{\rho}^{2\gamma/\text{EMD},\mathrm{X}}_{\mathrm{LCW}}(\vec{k}_\perp)
  =\!\!
  \sum_{\{{\bf K}_\perp\}}
  \!\!
  \tilde{\rho}^{2\gamma/\text{EMD},\mathrm{X}}({\bf p}_\perp)
  \Big\vert_{{\bf p}_\perp = {\bf k}_\perp + {\bf K}_\perp}
  \!\!,
\end{equation}
where the sum runs over a set of translation vectors in the projection plane, ${\{{\bf K}_\perp\}}$.
We analyzed the LCW back-folded radial anisotropy of EMD and $2D$-ACAR instead of commonly computed LCW back-folded EMD and $2D$-ACAR.
In this case the FS features are better visible and electron- and hole-like FS sheets better recognisable (cf. Figs.~\ref{fig:Fig5}-\ref{fig:Fig8} with Figs.~\ref{fig:Fig6_B}-\ref{fig:Fig8_B}). 
Unlike the ``high-pass'' like filters, where the Gaussian convoluted data is subtracted from the unconvoluted one, the order of computing the radial anisotropy and LCW back-folding is not interchangeable:
first the radial anisotropy should be computed and obtained result, $\tilde{\rho}^{2\gamma/\text{EMD},\mathrm{X}}_{\mathrm{an}}({\bf p}_\perp)$, back-folded with the LCW procedure~\eqref{eq:LCW}.
The radial anisotropy offers some advantage to high-pass filters, because the experimental $2D$-ACAR measurements contain both the core-electron and filled-band contributions which in general rather isotropic, and can be eliminated by this procedure. Moreover, positron annihilation experiments are sensitive to the presence of open-volume defects, and it has been shown that performing the back-folding procedure on the radial anisotropy can suppress their deleterious effect in the LCW densities \cite{dugdale2014}. 

The back-folded spectra of the radial-anisotropy is shown in the left plot in Fig.~\ref{fig:Fig5}. 
All Fermi surface sheets can be identified in the back-folded radial anisotropy plots (also {\it cf.} Fig.~\ref{fig:Fig4}):
large electron sheet around the $\Gamma$-point can be identified as red squares with centers at $\Gamma$-point;   
In the EMD, the corners of the projected $\Gamma$-centered electron Fermi surface sheets are``chopped'' (values are negative), while in the $2D$-ACAR the corners become positively weighted again.
one sees the small circle shaped deeps in the center of this red squares corresponding to two out of six face centered $X$-hole pockets (square faces of fcc BZ);
remaining four that overlap could be identified at $(X,W)$-points as dark-blue squares;
the ``jungle-gym'' (tubes together with $X$-hole fins) becomes also clearly visible in both EMD and $2D$-ACAR back-folded spectra (plus shapes with center at either $(\Gamma,X)$-points or $(X,W)$-points);
ellipsoid shaped $L$-hole pockets are well recognized in $2D$-ACAR in both LDA and LDA+DMFT cases as well es in EMD for LDA+DMFT.
Note that, redistribution of electronic momenta around the $L$-point due to the presence of electronic correlation or positrons does not alter the signal from the $L$-hole pockets and it should be visible in a high resolution $2D$-ACAR experiment e.g., using a spectrometer based on high-density avalanche chambers~\cite{Dugdale_2013}.
In Fig.~\ref{fig:Fig6} we show broadened $2D$-ACAR LDA+DMFT results for the relativistic ${c_0/c=1}$ and non-relativistic ${c_0/c=0}$ cases.
Comparing results for relativistic calculations, where we expect $L$-hole pockets in FS, with non-relativistic ones, which does not exhibit this feature, one can distinguish the $L$-hole pocket features with a momentum resolution FWHM of $=0.04$~a.u., which is about the width of the $L$-hole pocket (see Table~\ref{tab:Tab1}). Therefore we conclude that $2D$-ACAR measurements with a comparable or a better resolution will be able to resolve this FS feature directly.

\subsection{Momentum densities and electron-positron correlation functionals}
\label{subsec:EMD_ACAR}

The theory of the annihilation probability of a positron in a homogeneous electron gas has a long history in many-body physics~\cite{ca.ka.65,carb.67,ar.pa.79,ru.st.88,bo.sz.81}.
The electron-positron attraction leads to an increase of the electron density near the positron. It manifests itself in the annihilation characteristics and leads to a strongly increased total annihilation rate which  is  the  integral  of $\rho({\bf p})$  over  the  momentum {\bf p}. 
This effect is qualitatively well understood and is called ``enhancement''.


Since many decades, a central topic of theoretical positron physics is the inclusion of the enhancement effect into the theory of annihilating electron-positron pairs.
Whereas this effect is --- at least in principle --- well understood for positrons annihilating in a homogeneous electron gas \cite{kaha.63,bo.sz.81,carb.67}, the evaluation of Kahana's equation:
\begin{align}\label{eq:Kahana}
\begin{split}
&G_{{\rm ep}}(x,x';y,y')
=
G_{{\rm e}}(x,y)G_{{\rm p}}(x',y') \\
&\!+\!\!\frac{i}{\hbar}\!\!\int
\!\!{\mathrm d}z {\mathrm d}z'\!
G_{{\rm e}}(x,z)G_{{\rm p}}(x',z') V^{{\rm ep}}\!(z,z')
G_{{\rm ep}}(z,z';y,y')\!
\end{split}
\end{align}
in spatially inhomogenous electron gases is still an extremely laborious task. 
In Eq.~\eqref{eq:Kahana} we employ composite coordinates ${x\equiv ({\bf x},t_x)}$ etc.,
where $G_{{\rm e}}$ and $G_{{\rm p}}$ are the single-particle Green's functions for an electron and a positron, respectively, and $V^{{\rm ep}}$ describes the effective potential between the two fermions (electron and positron).
The simplest approximation for the electron-positron Green's function is given by the first term of Eq.~\eqref{eq:Kahana}: ${G_{{\rm ep}}(x,x';y,y') \approx  G_{{\rm e}}(x,y)\, G_{{\rm p}}(x',y')}$
and it ignores all terms which include the electron-positron potential $V^{ep}$.
This means that all direct correlation effects between the annihilating particles (leading to the so-called enhancement of the electron-positron annihilation rate) are neglected.
It represents the so-called independent particle model (IPM).
However, this does not mean that electron and positron interactions with their electronic environment are completely ``switched off'' in this approximation.
These interactions can be, at least, partly included into the theory by self-energy insertions into the one-particle propagators $G_e$ and $G_p$.

Using the product of the electron and positron Green's functions, the electron-positron momentum density ($\rho^{2\gamma}$)
can be written in terms of  
the Fourier transform of the overlap of Bloch functions of the annihilating electron and positron.
A frequently used non-ab initio approach for a description of $\rho^{2\gamma}$ including direct electron-positron correlations is based on the insertion of a local enhancement function, into the overlap of the electron and annihilating positron wave functions.
Such a procedure was first discussed by Carbotte~\cite{carb.67} and Fujiwara et al.~\cite{fu.hy.72}, and has been further developed by Sormann~\cite{sorm.96} under the name Bloch-modified ladder (BML) theory. 
The BML theory is quite successful in giving --- at least qualitatively --- a correct description of the electron-positron momentum density in metals of very different electronic structures~\cite{so.so.01,so.ko.06}, but on the other hand, this approach is numerically expensive and has some convergence problems with respect to the unoccupied electron bands.

For the electron momentum density in the IPM, there are no dynamical electron-electron or electron-positron correlations. 
Only in the Fourier integral correlations between the annihilating particles are approximately taken into account by the non-dynamical (static) enhancement function.
During the last decades, several proposals for this enhancement function have been made.
Here we shall mention briefly only the two most representative ones, namely:
{\em i)} formulas for a momentum-independent enhancement function, based on Lantto's work~\cite{lant.87} about spatially homogeneous electron-positron plasmas and presented by Boronski and Nieminen (see Ref.~\cite{bo.ni.86}) and by Puska {\it et al.}~\cite{pu.se.95}, and
{\em ii)} a  state-dependent enhancement factor published by Alatalo et al.~\cite{al.ba.96} and by Barbiellini et al.~\cite{ba.ha.97}. As has been already mentioned in connection with the electron-positron correlation potential, this enhancement factor can also be used either on an LDA level \cite{bo.ni.86,pu.se.95} or on a GGA level~\cite{ba.pu.95,ba.pu.96}.
This approach was generalized to an energy dependent form~\cite{mi.si.79} and was later extended to include orbital dependence~\cite{si.ma.86}.
Since this was formulated within DFT~\cite{bo.ni.86,ru.st.88,ba.pu.95,ba.pu.96,Barb97,sorm.96,so.ko.06,ko.so.12,ko.so.14,Kontrym-Sznajd2014,boro.14} the results maintain their static mean-field character, but now they also reflect the inhomogeneity of the electron gas of real materials.
Sormann and Kontrym-Sznajd investigated the effectiveness of various enhancement models applied to different metals like Mg, Cd, Cu, and Y.
They compared the theoretical MDAP distributions with experimental data and with the ab-initio approach BML (see Ref.~\cite{so.ko.02}).
The main result of this work was that all investigated non-ab initio enhancement models have their advantages and disadvantages, i.e. they succeed for one metal and fail for another.
Nevertheless, it should be emphasized here that all theoretical approaches mentioned above (including BML) describe a static enhancement without any dynamical effects.

\begin{figure}[!thbp]
  \includegraphics[width=0.99\linewidth]{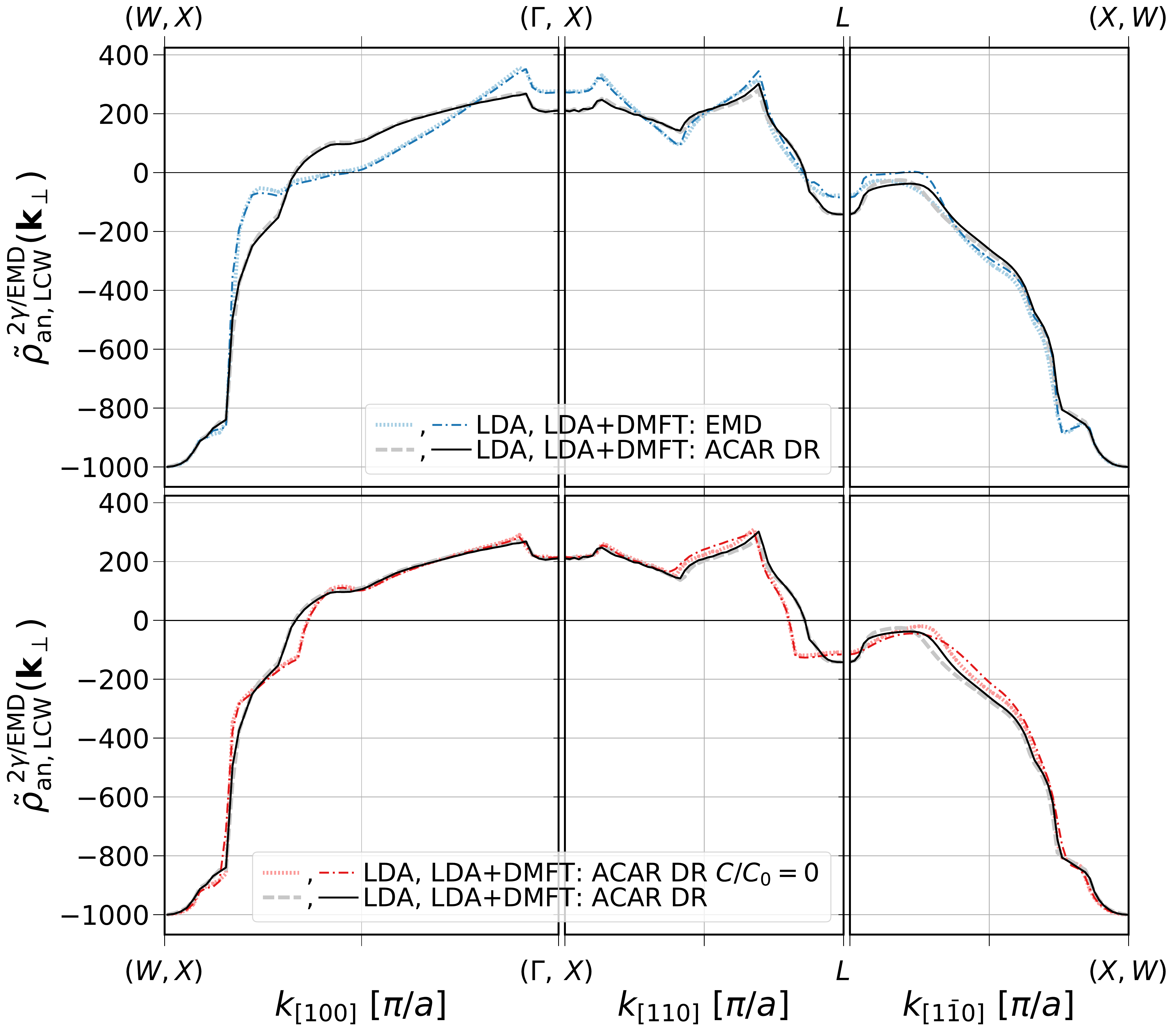}
  \caption{Closed path ${X-(\Gamma,X)-L-X}$ (triangle) across the LCW back-folded data. Upper row corresponds to the results presented in Fig.~\ref{fig:Fig5}
          }
  \label{fig:Fig7}
\end{figure}

To model the enhancement we considered the formulation proposed by Drummond {\it et al.}~\cite{dr.lo.11}. 
In Fig.~\ref{fig:Fig7} we show sections of the back-folded momentum densities through the $L$-hole and $X$-hole pockets.
The general shape of the $2D$-ACAR and the EMD is similar, and the structure seen in these curves can be related to the Fermi-surface topology and the nature of the wave functions.
One clearly sees deeps at $(\Gamma,X)$- and $(W,X)$-points where the projected $X$-hole pockets are expected and one can also read out the linear size of the ellipsoid of revolution.
At about a half of the $(W,X)-(\Gamma,X)$ distance, where one still expects the contributions from the $\Gamma$-centered electron surface, the calculated weight dives below $0$ in the EMD, whereas in the $2D$-ACAR spectra one can recognize the edge of this large electron Fermi surface.
We note that the $L$-hole pocket has a similar width in the EMD and $2D$-ACAR spectra, while it is deeper in the case of the $2D$-ACAR spectra in which enhancement of momentum density is visible. 
Further on one sees that the relativistic effects are the determining factor in the appearance of the $L$-hope pocket.
In the lower panel of Fig.~\ref{fig:Fig7} where we show the comparison of the relativistic versus the non-relativistic ($c/c_0=0$) $2D$-ACAR spectra one sees that the non-relativistic spectra are flattened where the $L$-hole pocket is expected in comparison with the relativistic case.
Whereas the linear sizes of the $X$-hole pocket remain almost similar in the non-relativistic ($c/c_0=0$) case (see the lower panel of Fig.~\ref{fig:Fig7}).

\begin{figure*}
  \includegraphics[width=0.49\linewidth]{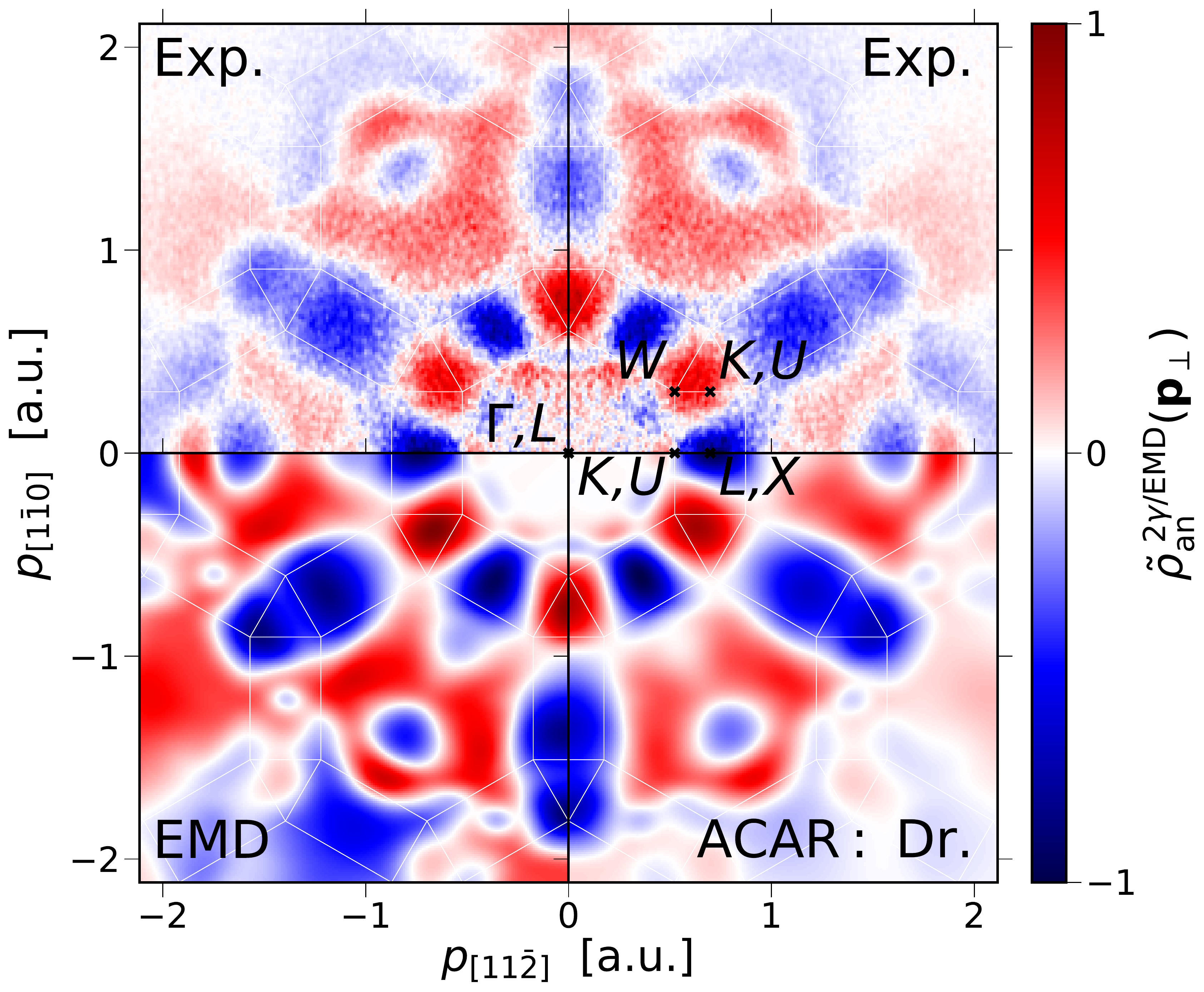}
  \hfill
  \includegraphics[width=0.49\linewidth]{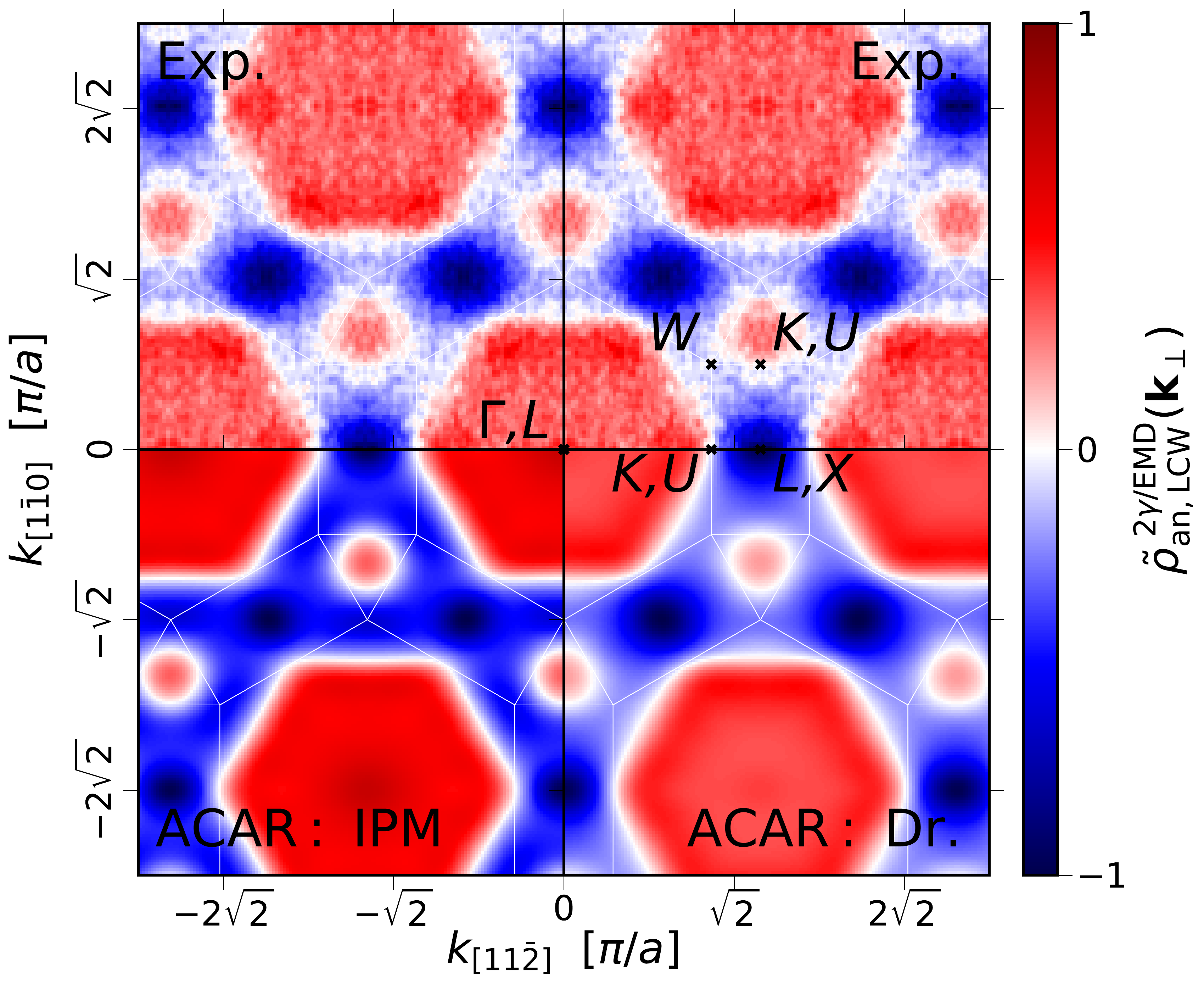}
  \caption{Anisotropy (left) and LCW back-folded anisotropy (right) plots of $2D$-ACAR spectra along $[111]$ high-symmetry direction of experimental measurements (upper rows) and LDA+DMFT calculations (${U=1.0}$~eV, ${J=0.3}$~eV) (lower rows). The EMD (lower-left quadrants). The Drummond parametrization of the enhancement factor is used for the $2D$-ACAR spectra (bottom-right quadrants).
           The false-color map is normalized by maximal amplitude of anisotropy in each cases.
          }
  \label{fig:Fig8}
\end{figure*}

\subsection{Comparison to the experimental measurement}
\label{subsec:EMD_EXP}

As already discussed in the Sec.~\ref{sec:band_FS}, the $2D$-ACAR spectra along $[110]$ direction will be beneficial in observing the $L$-hole pockets, since in this direction other Fermi surface sheets/pockets do not overlap with the $L$-hole pocket in the projected spectra.
Nonetheless in absence of recent accurate measurements, 
on high quality bulk single crystal of Pd along the $[110]$ direction,
we analyze the accuracy of our calculation and draw conclusions about the ability of simulations to reproduce/predict the observed structure. 
This projection however overlays contributions from around the $L$-point to those from the central $\Gamma$-point and the larger size $X$-hole pockets, and therefore does not provide unambiguous information about $L$-pockets.

A $2D$-ACAR experiment was performed on a bulk single-crystal of Pd using a spectrometer at the University of Bristol, the design of which is described in Ref.~\cite{West1981}. The measurement was made at a temperature of 30K.
The angular resolution full widths at half-maximum (FWHM) was $\approx 0.09$~a.u. and $\approx 0.14 $~a.u. along the $[1\bar{1}0]$ and $[11\bar{2}]$ directions, respectively.

In order to expose details of the momentum densities the anisotropy spectrum $\rho_{an}({\bf p}_\perp)$ is obtained by subtracting an angular averaged contribution $\overline{\rho}(p)$. This procedure eliminates radially symmetric contributions from core electrons and in metals can be used to highlight the Fermi surface features.

In Fig.~\ref{fig:Fig8} we present a direct comparison of the experimental measurement and theoretical calculations (including electronic correlations) for the anisotropy spectra  in the extended momentum scheme (left plot) and the corresponding LCW back-folded spectra (right plot).
We observe the very good agreement between the experimental and the computed $2D$-ACAR spectra which includes the Drummond's~\cite{dr.lo.11} enhancement scheme. 
We have also included the EMD spectra.
The theoretical distributions were convoluted with a Gaussian with $0.097$~a.u. and $0.137$~a.u. FWHM presenting the experimental resolution in $[110]$ and $[112]$ directions, respectively.
Characteristic to the experimental spectra for this particular projection is the (six fold symmetry) hexagonal pattern of bright and dark areas which is very well captured by the $2D$-ACAR calculation.   
Although in several momentum regions the departure with respect to the experiment is clearly seen, overall, the inclusion of electron-positron enhancement effects increases the intensity of the anisotropy signal for small momenta, and improves the comparison with the experiment.

On the right hand side of Fig.~\ref{fig:Fig8} we show the LCW back-folded spectra. Around the $\Gamma$-point electronic momentum of considerable weight is found within a hexagonal shaped region. The $2D$-ACAR computation reproduces very well the size and the shape of the hexagonal region. For the EMD spectra we see that the hexagon's boundary (sides) are  somewhat reduced in size and are slightly concave.
Outside the central region inside the triangles, produce by overlapping of the projections of neighboring BZs, small high positive intensity regions are seen following the hexagonal pattern.
There are also ellipse-shaped negative intensity features due to $X$- and $L$-hole pockets ($X$-hole pocket projection completely covering the $L$-hole pocket projection) organized in the hexagonal pattern around the central dominant high intensity convex hexagon with its center at $\Gamma$.
All these features are also well reproduced by the theoretical $2D$-ACAR calculations including positron wave function and the electron-positron interaction enhancement factor parametrized by Drummond~\cite{dr.lo.11}.

In a recent paper~\cite{ke.bi.21} the differences between the momentum distributions observed by positron annihilation and Compton scattering were investigated.
The focus of that study was a new approach for tomographical reconstruction of $2D$ electron momentum distributions from the $1D$ Compton profiles rather than observing the impact of electron correlation and relativistic effects on electronic structure of Pd and its Fermi surface.
The experimental resolution of $2D$-ACAR spectra presented in the current paper ($0.097$~a.u. and $0.137$~a.u.), however, is significantly better than the values reported in Ref.~\cite{ke.bi.21} ($0.17$ and $0.21$~a.u.).
The inferior experiment resolution reported in Ref.~\cite{ke.bi.21} is due to the spectrometer's relatively short detector-detector distance.  

%

\section{Summary}
\label{sec:SUMM}

The primary purpose of the present computations was to investigate to what extent electronic correlations may affect the Fermi surface of Pd.
The main result is that band theory including relativistic effects accounts relatively well for the experimentally observed features of the Fermi surface, in particular the existence of small $L$-hole pockets. 
Within the current advances of realistic many-body computations we have performed LDA+DMFT electronic structure calculation and discussed the modifications in the band structure and Fermi surface subject to the change in the strength of relativistic effects. 

The Fermi surface of Pd has been explored by a variety of different methodologies such as angle-resolved photoemission (ARPES), de Haas-van Alphen (dHvA) and ultrasonic attenuation, which finally established the existence of $L$-hole pockets not in a direct way, though.
Complementary to these methods we discuss here the two-dimensional angular correlation of annihilation radiation ($2D$-ACAR) positron spectroscopy which directly probes the electronic structure by mapping out the Fermi surface and enables the observation of fine structures such as the $L$-hole pockets in discussion. 
Since the positron tends to occupy open volumes as it tries to avoid positively charged ionic cores, the positron spatial distribution in a crystal is in general quite non-uniform. In the specific case of Pd annihilation events happen between the positrons and Bloch state electrons of the compact fcc crystal structure thus, one also expects that the enhancement (electron-positron interaction effects) might be essential. 
A comparison of the experimental and theoretical $2D$-ACAR spectra (momentum densities in the extended and back-folded schemes) has been done for the [111] projection. The comparison shows that there is a reasonable agreement to the overall shape as well as to the fine structure in the spectra. The importance of enhancement effects becomes visible by contrasting the $2D$-ACAR with the electron momentum densities. By comparing different functionals (results not shown) for the electron positron interaction we found the Drummond's parametrization captures the experimental features the best. 
Finally we note that, in terms of electronic correlations, for the strength of the Coulomb and Hund parameters ($U$,$J$) that properly describe the ground state properties in Pd the changes in the $2D$-ACAR spectra are mild.

Because of the observed good agreement between the calculated and measured $2D$-ACAR spectra it is natural to expect that $2D$-ACAR measurements along $[110]$ direction with a resolution higher than or about $0.04$~a.u. will be able to reveal the $L$-hole pockets structure in Pd Fermi surface topology.

\acknowledgements

Financial support by the Deutsche Forschungsgemeinschaft through TRR80 (project E2) Project number
107745057 is gratefully acknowledged.
M.S.~was partially supported by the Shota Rustaveli Georgian National Science Foundation through the grant N~FR-19-11872.
We would like to thank J.~Min\`ar and H.~Ebert for a fruitful collaboration.

\appendix*{}
\section{LCW back-folded data}

\begin{figure}[thbp]
  \includegraphics[width=0.99\linewidth]{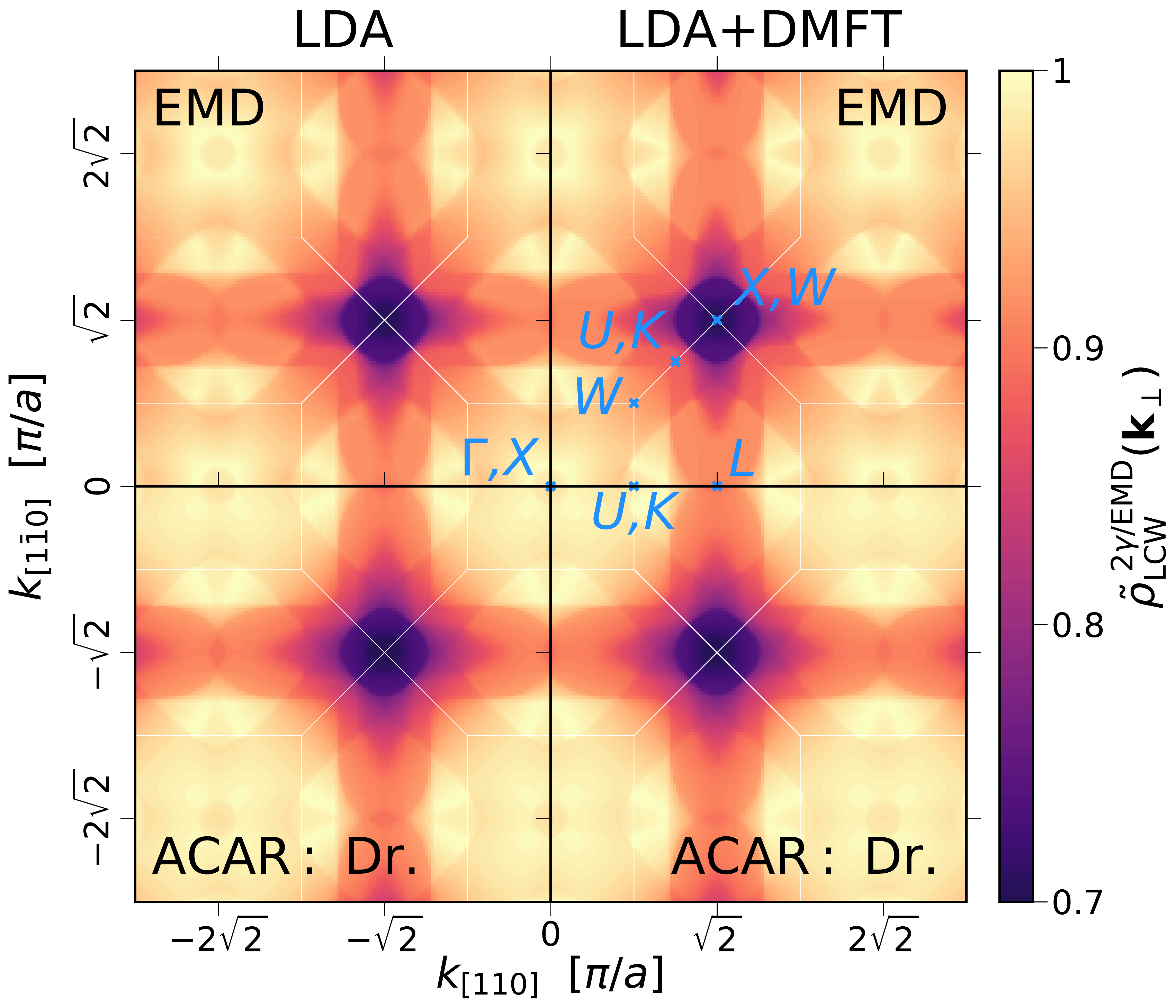}
  \caption{LCW back-folded $2D$ projection of EMD (upper row) and $2D$-ACAR spectra (bottom row) from LDA (left column) and LDA+DMFT (${U=1.0}$~eV, ${J=0.3}$~eV) (right column) calculations.
           The false-color map is normalized for each case by the maximum value of ${\max[\rho_{\mathrm{LCW}}(\mathbf{k}_\perp)]}$.
           Projections of the high-symmetry points (cyan crosses), as well as extended BZ (white lines) are also shown.
           The corresponding LCW back-folded radial anisotropy is shown in the right plot in Fig.~\ref{fig:Fig5}.
          }
  \label{fig:Fig6_B}
\end{figure}

\begin{figure}[!th]
  \includegraphics[width=0.99\linewidth]{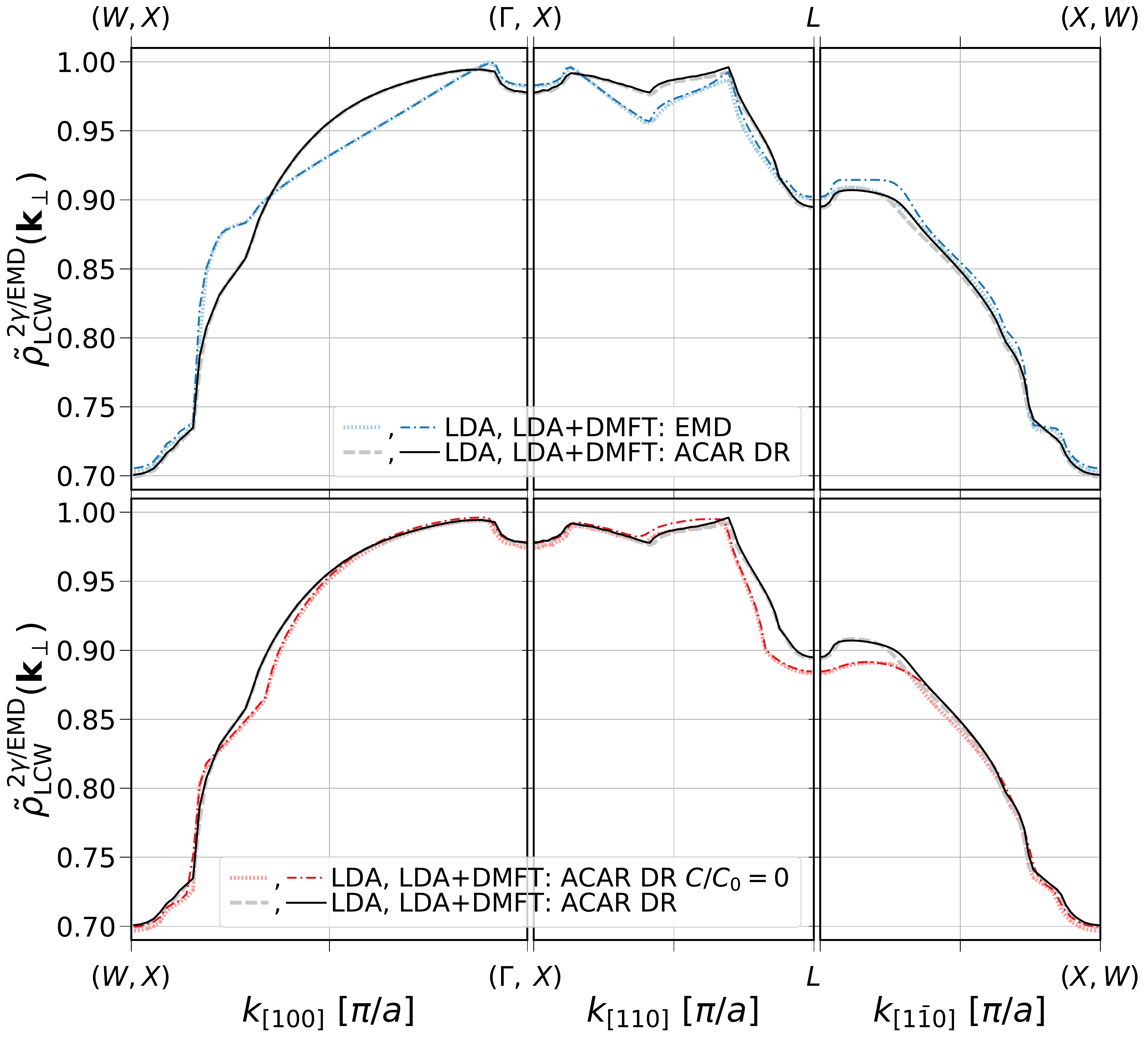}
  \caption{Closed path ${X-(\Gamma,X)-L-X}$ (triangle) across the LCW back-folded data, corresponding to the calculations presented on Fig.~\ref{fig:Fig6_B}.
           The corresponding radial anisotropy is shown in Fig.~\ref{fig:Fig7}.
          }
  \label{fig:Fig7_B}
\end{figure}

\begin{figure}[!th]
  \includegraphics[width=0.99\linewidth]{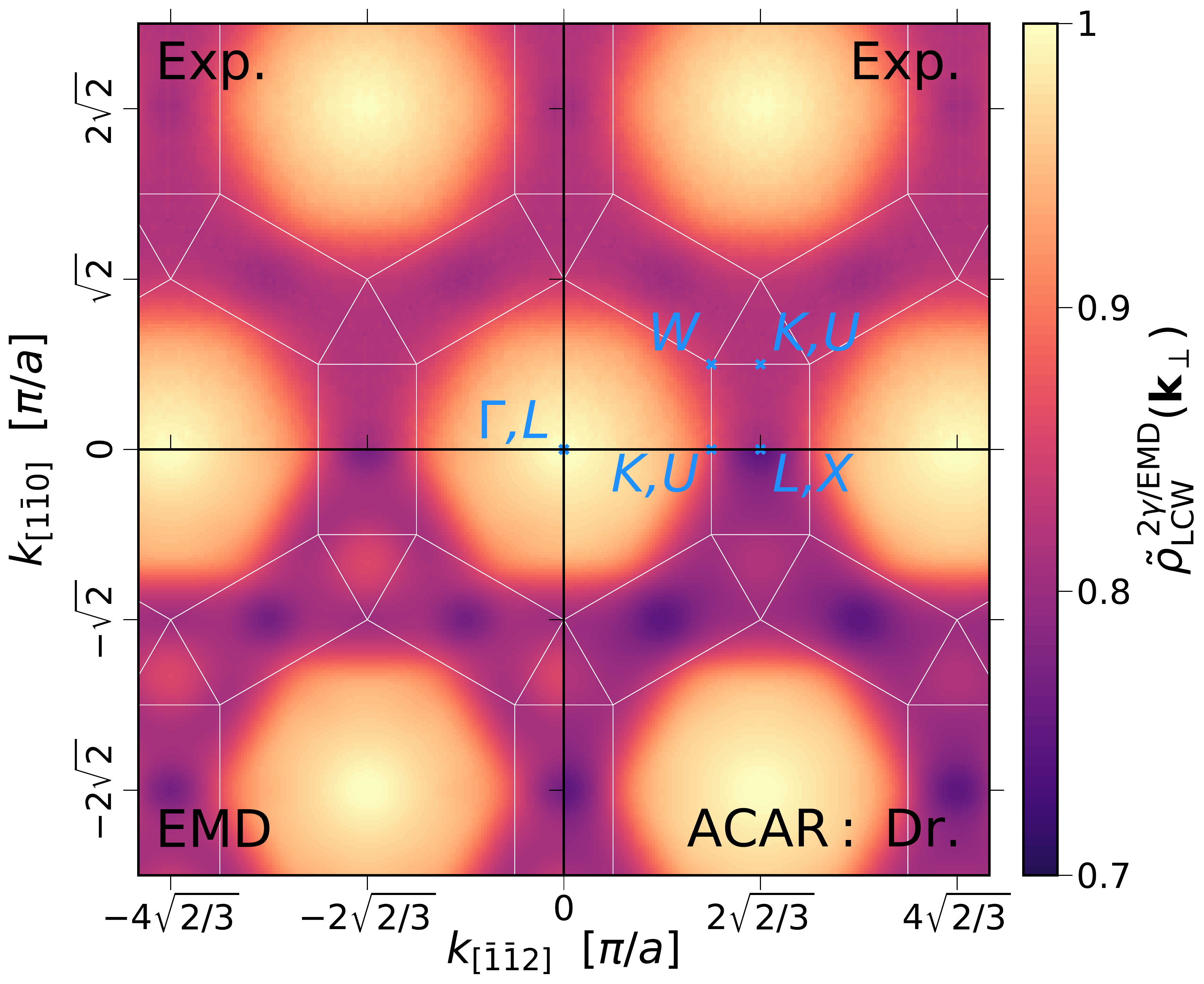}
  \caption{LCW back-folded $2D$-ACAR spectra along $[111]$ high-symmetry direction of experimental measurements (upper row) and LDA+DMFT calculations (${U=1.0}$~eV, ${J=0.3}$~eV) (lower row).
           The false-color map is normalized for each case by the maximum value of ${\max[\tilde{\rho}_{\mathrm{LCW}}(\mathbf{k}_\perp)]}$.
           Projections of the high-symmetry points (cyan crosses), as well as extended BZ (white lines) are also shown.
           The corresponding LCW back-folded radial anisotropy is shown in Fig.~\ref{fig:Fig8}.
          }
  \label{fig:Fig8_B}
\end{figure}

For the sake of completeness in Fig.~\ref{fig:Fig6_B} and Fig.~\ref{fig:Fig8_B} we show the LCW back-folded EMD and $2D$-ACAR results corresponding to the radial anisotropies shown in the right plots in Fig.~\ref{fig:Fig5} and Fig.~\ref{fig:Fig8}, respectively.
The former is typically reported in the publications, but we find that LCW back-folded radial anisotropy better highlights electron and hole like features of the Fermi surface sheets. As a counter part of the data shown in Fig.~\ref{fig:Fig7}, in Fig.~\ref{fig:Fig7_B} we show the sections of the LCW back-folded EMD and $2D$-ACAR (sections of the data shown in Fig.~\ref{fig:Fig6_B}) (upper panel) together with non-relativistic results.
One can also see the $L$-pocket structure in the relativistic calculation. Note it's absence in the non-relativistic case, as similarly can be seen in the cuts of LCW back-folded anisotropy (see Fig.~\ref{fig:Fig7}).

\clearpage

\bibliography{main}

\end{document}